\documentclass{PoS}

\usepackage{amsmath}
\title{RHMC simulation of two-dimensional N=(2,2) super Yang-Mills with exact supersymmetry }

\ShortTitle{RHMC simulation of two-dimensional N=(2,2) super Yang-Mills with exact supersymmetry }

\author{\speaker{Issaku Kanamori}\\%
        Theoretical Physics Laboratory, RIKEN, 2-1 Hirosawa, Wako,
         Saitama 351-0198, Japan\\
        E-mail: \email{kanamori-i@riken.jp}}

\abstract{
We report our numerical simulation of two-dimensional $\mathcal{N}=(2,2)$
super Yang-Mills.  The lattice model we use is one proposed by F.~Sugino
which keeps one exact supersymmetry at finite lattice spacing.  
We use Rational Hybrid Monte Carlo (RHMC) method to implement 
the dynamical fermion.
We apply the simulation to measure the ground state energy which is
useful to observe dynamical SUSY breaking.
}

\FullConference{The XXVI International Symposium on Lattice Field Theory \\
		 July 14 - 19, 2008\\
		 Williamsburg, Virginia, USA\\
Report Number: RIKEN-TH-138}

\newcommand{\phibar}{\overline{\phi}}
\newcommand{\tr}{\mathop{\rm tr}\nolimits}

\begin{document}

\section{Introduction}

It seems impossible to put the SUSY on the lattice, because
SUSY algebra contains infinitesimal translation but on the lattice
we have only finite translations.
However, what we have realized in the recent development is that
it is possible to formulate supersymmetric models on the lattice 
if $\mathcal{N}\geq 2$.
Lots of lattice models especially for 
super Yang-Mills are known along this line 
\cite{Kaplan:2002wv, Sugino:2003yb,Catterall:2004np,Suzuki:2005dx,D'Adda:2005zk}
and relations among them have become clear \cite{
Takimi:2007nn,Damgaard:2007xi,Damgaard:2007eh}.
\footnote{
For the review, see \cite{Giedt:2007hz}. 
} 
Most of these formulations utilize the topological twist.
After the twist, we have a scalar supercharge instead of spinors.
We can put the scalar on a lattice site and keep it exactly at finite
lattice spacing.\footnote{
For recent developments in keeping whole supersymmetry exactly on the
lattice,  see
\cite{Nagata:2008xk,Arianos:2008ai}; See also \cite{Kato:2008sp}.
}
Some of the simulation have already done aiming the check
of the formulation
\cite{Catterall:2006jw,Suzuki:2007jt,Fukaya:2007ci}.\footnote{
See \cite{Elliott:2005bd,Elliott:2007bt,Elliott:2008jp}
 for interesting attempts towards simulation in three and
four dimensions.
}

In this talk we report our simulation with dynamical fermions and its 
application.  
We utilize the Rational Hybrid Monte Carlo algorithm \cite{Clark:2004cp}.
The target model is two-dimensional $\mathcal{N}=(2,2)$ super Yang-Mills model
based on a formulation with one exactly kept supersymmetry proposed by 
Sugino \cite{Sugino:2003yb}.
As an application of the simulation, we measure the ground state energy
which is useful to observe dynamical SUSY breaking.
We also sketch the method of observing dynamical SUSY breaking which 
we proposed in \cite{prd,ptp}.

\section{Lattice Model}
The target theory in the continuum has $\mathcal{N}=(2,2)$ twisted 
supersymmetry.
After the twist we have four supercharges, one from a scalar $Q$, two from
a two-dimensional vector $(Q_0,\ Q_1)$, and one from a pseudo scalar
$\tilde{Q}$.
The following is a part of the twisted SUSY algebra:
 \begin{align}
  Q^2 &= \delta_\phi^{\rm (gauge)},
  &
  Q_0^2 &= \delta_{\phibar}^{\rm (gauge)}, &
 \{Q, Q_0\} &= 2i\partial_0 +2\delta_{A_0}^{\rm (gauge)},
  \label{eq:algebra}
 \end{align}
where $\delta_{\ \bullet}^{\rm (gauge)}$ denotes an infinitesimal gauge
transformation with the parameter $\bullet$.
The supercharges are nilpotent up to gauge transformation.
The action is $Q$-exact and because of the nilpotency, $Q$-invariance is
manifest. 

On the lattice, we keep the scalar $Q$ exactly.
The following $Q$ transformation on the lattice 
keeps the nilpotency even at the finite lattice spacing \cite{Sugino:2003yb}:
\begin{align*}
 Q U(x,\mu) &= i\psi_\mu(x) U(x,\mu), 
&  Q \psi_\mu(x)&=i\psi_\mu(x)\psi_\mu 
      -i\bigl(\phi(x)-U(x,\mu)\phi(x+\hat{\mu}\bigr)U(x,\mu)^{-1}), \\
 Q \phi(x) &= 0,\\
 Q \chi(x) &= H(X),
& Q H(x) &= [\phi(x), \chi(x)], \\
 Q \phibar(x) &= \eta(x),
& Q \eta(x) &= [\phi(x), \phibar(x)],
\end{align*}
where $U(x,\mu)$ is a gauge link variable, 
scalar fields $\phi, \phibar$ and auxiliary field $H$ are defined on the sites,
fermions in the twisted basis $\eta, \chi, \psi_\mu$ are defined 
on the sites.
The action is defined as $Q$-exact as in the continuum case:
 \begin{align*}
  S&= Q \frac{1}{a^2g^2} \sum_x \tr\biggl[
      \chi(x)H(x)
      +\frac{1}{4}\eta(x)[\phi(x), \phibar(x)]  
     -i\chi(x) \hat{\Phi}(x)  \\ 
  &\quad  + i\sum_{\mu=0,1}\left\{\psi_\mu(x)\left(\phibar(x)
  -U(x,\mu)\phibar(x+a\hat{\mu})U(x,\mu)^{-1}\right)\right\} 
\biggr] \\
  &= \frac{1}{a^2g^2}\sum_x \tr \biggl[
     \frac{1}{4}\hat{\Phi}_{\rm TL}(x)^2 + \dots \qquad
\biggr],
 \end{align*}
where $g$ is the dimensionful gauge coupling,
$i\hat{\Phi}(x) = \dfrac{U(x,0,1)-U(x,0,1)^{-1}}{1- 
||1-U(x,0,1)||^2/\epsilon^2 }$, 
$\hat{\Phi}_{\rm TL}$ is the traceless part of $\hat{\Phi}$ and
$U(x,0,1)$ is the usual plaquette variable.
We impose the admissibility condition $ ||1-U(x,0,1)|| <\epsilon $ 
for a constant $\epsilon$ in order to kill artifact vacua. 
Because of the nilpotency and $Q$-exactness, the action is manifestly 
$Q$ invariant at the finite lattice spacing.
The other three supercharges, $Q_0$, $Q_1$ and $\tilde{Q}$, will be 
automatically restored in the continuum limit as long as a perturbative
power counting is valid.

\section{Simulation Details}
Since fermions play an important role in supersymmetry the effect of the 
dynamical fermion is crucial.
We use the Rational Hybrid Monte Carlo algorithm.
The path integration of the fermion gives Pfaffian of the Dirac
operator $D$, which contains the Yukawa interaction terms as well.
We rewrite the Pfaffian using pseudo fermion integration
with rational function.  Symbolically, contribution from the fermionic
part of the action $S_{\rm fermion}$ becomes
\begin{align}
         \int \mathcal{D}f \exp(-S_{\rm fermion})
        &= \mathrm{Pf}(D)
         = \int \mathcal{D}F \exp(-F^\dagger (D^\dagger D)^{-1/4} F) \nonumber\\
        &= \int \mathcal{D}F \exp\left(-F^\dagger 
      \left[ a_0 + \sum_{i=1}^n \frac{a_i}{D^\dagger D +b_i}\right]F\right),
\end{align}
where $f$ is the fermion, $F$ the pseudo fermion, $D$ the Dirac
operator, $a_i$ and $b_i$ are numerical constants.\footnote{
We use a program from \cite{remez} to obtain $a_i$ and $b_i$.
}
Here we ignore a phase factor of the Pfaffian $\mathrm{Pf}(D)$ 
because it is almost $1$ (i.e., almost real and positive) 
in the current model.  If it is needed we would 
reweight this phase factor in the measurements.
We also utilize the multi time step evolution in the molecular dynamics 
\cite{Sexton:1992nu}.  
We calculate forces from $S_{\rm fermion}$ every several calculations of forces
from the other part of the action.  We evolve pseudo fermion $F$ as well as
other bosonic fields $U(x,\mu)$, $\phi$ and $\phibar$ 
in the molecular dynamics.

The parameters we use are the following.
We set the gauge group to $SU(2)$ and $\epsilon$ for the admissibility
condition to $2.6$.\footnote{
The possible maximum value for $\epsilon$ is $2\sqrt{2}=2.8284...$ in $SU(2)$
case.
}
The lattice size is $3\times 12$--$30\times 10$ and the lattice
spacing is $ag=0.07071$--$0.2357$.  
The degree of the rational approximation is typically $20$.
The length of time evolution in the molecular dynamics in each
trajectory is fixed to $0.5$.
We store the configurations every 10 trajectories.
We keep the acceptance in the Metropolis test to be greater 
than $0.8$.  Because of the flat directions in the scalar potential which
will be discussed later, and since 
the magnitude of the molecular dynamical 
force depends on the magnitude of the scalar fields,
the acceptance fluctuates during the simulation.\footnote{
This is for the case in which we impose anti-periodic boundary condition
in time direction for fermions.  In the periodic case, the flat
direction is lifted so that we do not observe such fluctuations of the 
acceptance.
}
The magnitude of the scalar fields tends to be larger and the 
acceptance tends to be smaller as the simulation runs.
See also Table~\ref{table:configurations} for the parameters and number
of configurations we use in the application.

\section{Application: observing dynamical SUSY breaking}
What can we do with this simulation?
We use it to observe the dynamical SUSY breaking using a method
we proposed in \cite{prd,ptp}.
The requirement for the lattice model in the method 
is that it should have nilpotent $Q$ and exact $Q$-invariance, which the
current model satisfies.
Since the SUSY is not broken in the perturbation if it is not broken in the
tree level, a way of observing SUSY breaking due to non-perturbative
effects is very important.
Usually, the Witten index provides such a method but in this system,
two-dimensional $\mathcal{N}=(2,2)$ pure super Yang-Mills, it is not available.
What we know without numerical simulations is an argument by Hori and 
Tong that SUSY is probably 
spontaneously broken in this system \cite{Hori:2006dk}.

What we measure is the ground state energy using 
the Hamiltonian as the order parameter.
As well known, vacuum expectation value of the Hamiltonian 
is zero if and only if the SUSY is not broken.
The advantage of using the Hamiltonian is that it requires one-point
function which is numerically much less expensive than two-point function.

Since we are interested in that it is zero or not,
the correct choice of the origin of the Hamiltonian is crucial.
We use the SUSY algebra to define the Hamiltonian.
We regard the anti-commutator in (\ref{eq:algebra}) as follow:
\begin{equation}
  Q \mathcal{J}_0^{(0)} =2\mathcal{H},
\end{equation} 
that is,
$Q$ transformation of the $0$-th component of the Noether current 
corresponding $Q_0$, $\mathcal{J}_0^{(0)}$, gives
the Hamiltonian density $\mathcal{H}$.
On the lattice we have only $Q$ transformation but no $Q_0$
transformation.
Therefore, we discretize the continuum version of the Noether current by
hand.
We use the following as the $0$-th component of the current for $Q_0$:
\begin{align}
 \mathcal{J}_0^{(0)} (x)
  &= \frac{1}{a^4g^2}\tr\biggl\{
     \eta(x)[\phi(x),\phibar(x)]^2 + 2\chi(x)H(x) \nonumber\\
  &\qquad -2i\psi_0(x)\bigl( \phibar(x)-U(x,0)\phibar(x+a\hat{0})U(x,0)^{-1}\bigr) 
  \nonumber\\
 &\qquad 
+2i\psi_1(x)\bigl( \phibar(x)-U(x,1)\phibar(x+a\hat{1})U(x,1)^{-1}\bigr)
\biggr\}.
\end{align}
Since we know the $Q$ transformation on the lattice 
so it is straightforward to obtain the Hamiltonian.

Another point is the boundary condition.
As usual method for observing spontaneous symmetry breaking,
we apply an external field conjugate to the order parameter.
The conjugate to the Hamiltonian is the temperature.
That is, we impose the anti-periodic condition in the time direction
for fermion.
Therefore we break SUSY by boundary condition or equivalently by the 
temperature.
Then we take zero temperature limit and see the effect of the 
breaking is left or not.

\begin{figure}
 \hfil
 \begin{minipage}[b]{.47\linewidth}
\includegraphics[width=.95\linewidth]{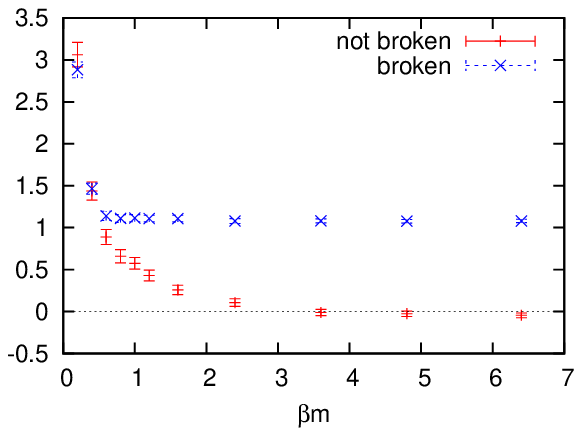}  
  \caption{Expectation value of the Hamiltonian for supersymmetric
  quantum mechanics versus inverse temperature $\beta$.  All quantities 
  are measured in a dimensionful parameter $m$ in the potential.}
  \label{fig:SQM}
 \end{minipage}
 \hfil
 \begin{minipage}[b]{.47\linewidth}
  \includegraphics[height=.95\linewidth,angle=-90]{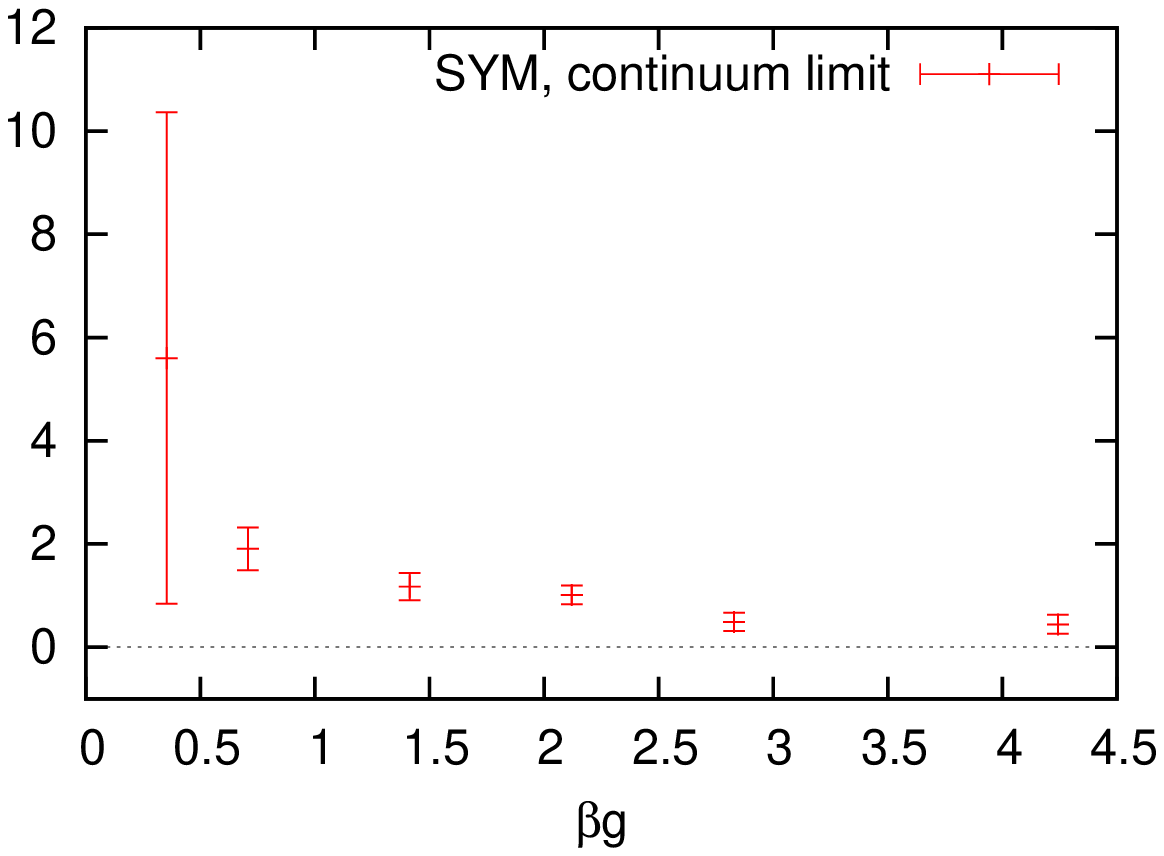}
  \caption{Expectation value of the Hamiltonian density for super
  Yang-Mills versus inverse temperature $\beta$.  All quantities 
  are measured in unit of dimensionful gauge coupling $g$.}
  \label{fig:SYM}
 \end{minipage}
\end{figure}

As a check of the method, we first investigate supersymmetric quantum
mechanics.
The known fact is that the form of potential decides whether 
SUSY is broken or not.
We use a lattice model given in \cite{Catterall:2000rv}, 
which has nilpotent $Q$
and $Q$-exact action.  The details of the measurement and forms of the
potential are found in \cite{ptp}.  
Here we only show Figure~\ref{fig:SQM}, from which 
we can easily distinguish SUSY broken case and not-broken case.
Our method actually works as expected.

Next let us investigate the super Yang-Mills case.
Figure~\ref{fig:SYM} shows the result.
We put the statistical errors only in the plot.
Although we can not deny a possibility of non-zero small ground state energy
that means breaking of the SUSY, 
the plot shows the value of the ground state energy is small and close
to zero.
We need to take a limit of the inverse temperature $\beta\to\infty$ but
the plot implies that the $\beta$ we use is enough large since the
expectation value of the Hamiltonian density is almost saturated.
Note that all quantities are measured in the dimensionful gauge coupling
$g$.
Some details of the measurement is in order.
We fix the physical spacial size $L_{\rm S} =1.41/g$.
We discard first 20,000--30,000 trajectories as thermalization.
We calculate the Hamiltonian every $10$ trajectories. 
In order to reduce the errors, we take an average over the lattice.
The errors are obtained using a jackknife analysis with binning.  
The bin size with which the autocorrelation disappears is 
typically $10$--$20$.  We list the number of the configurations
 after the binning in Table~\ref{table:configurations}.

\begin{figure}
 \hfil
 \begin{minipage}[t]{.47\linewidth}
  \includegraphics[height=.95\linewidth,angle=-90]{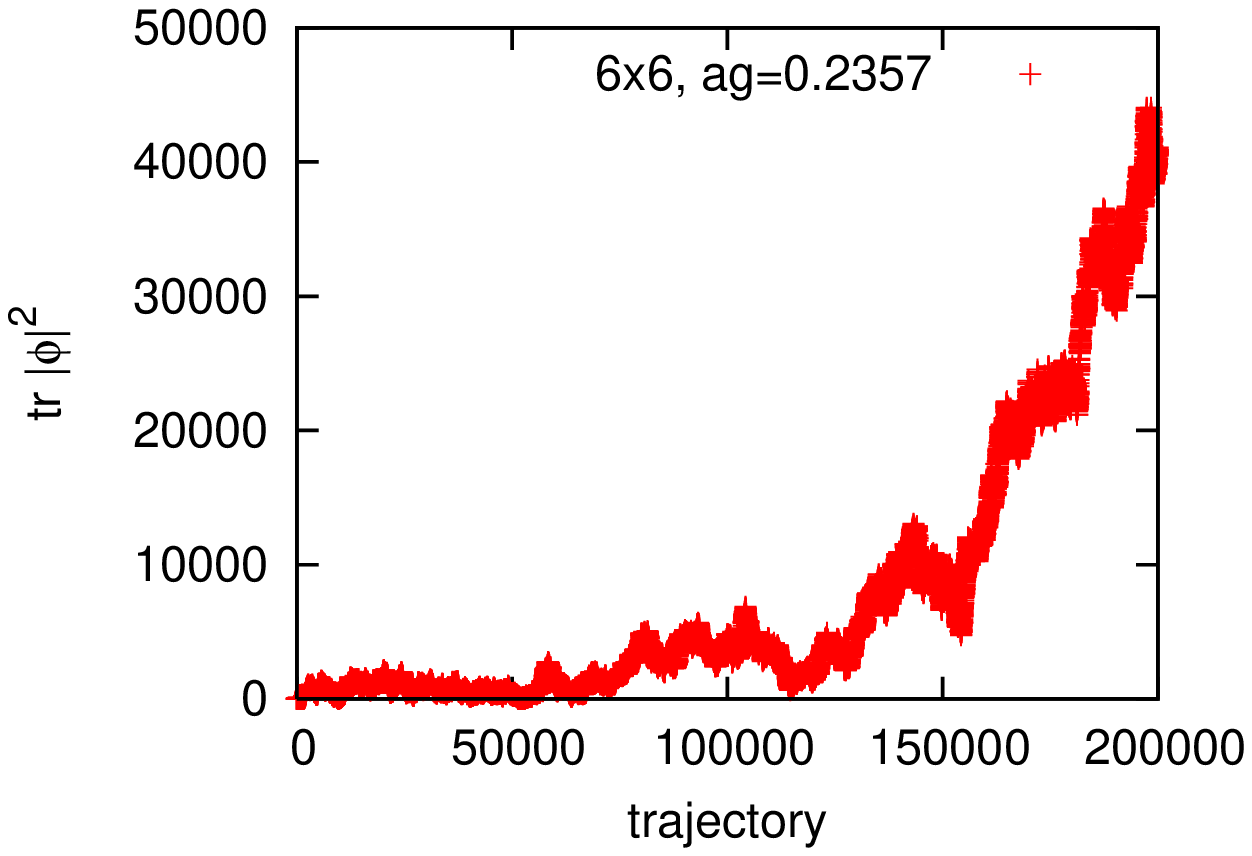}
  \caption{Evolution of the scalar norm over the trajectories in the
  anti-periodic case. The scalar norm is measured by $1/g^2$.
  It tends to be larger as the simulation runs.}
  \label{fig:scalar}
 \end{minipage}
 \hfil
 \begin{minipage}[t]{.47\linewidth}
  \includegraphics[height=.95\linewidth,angle=-90]{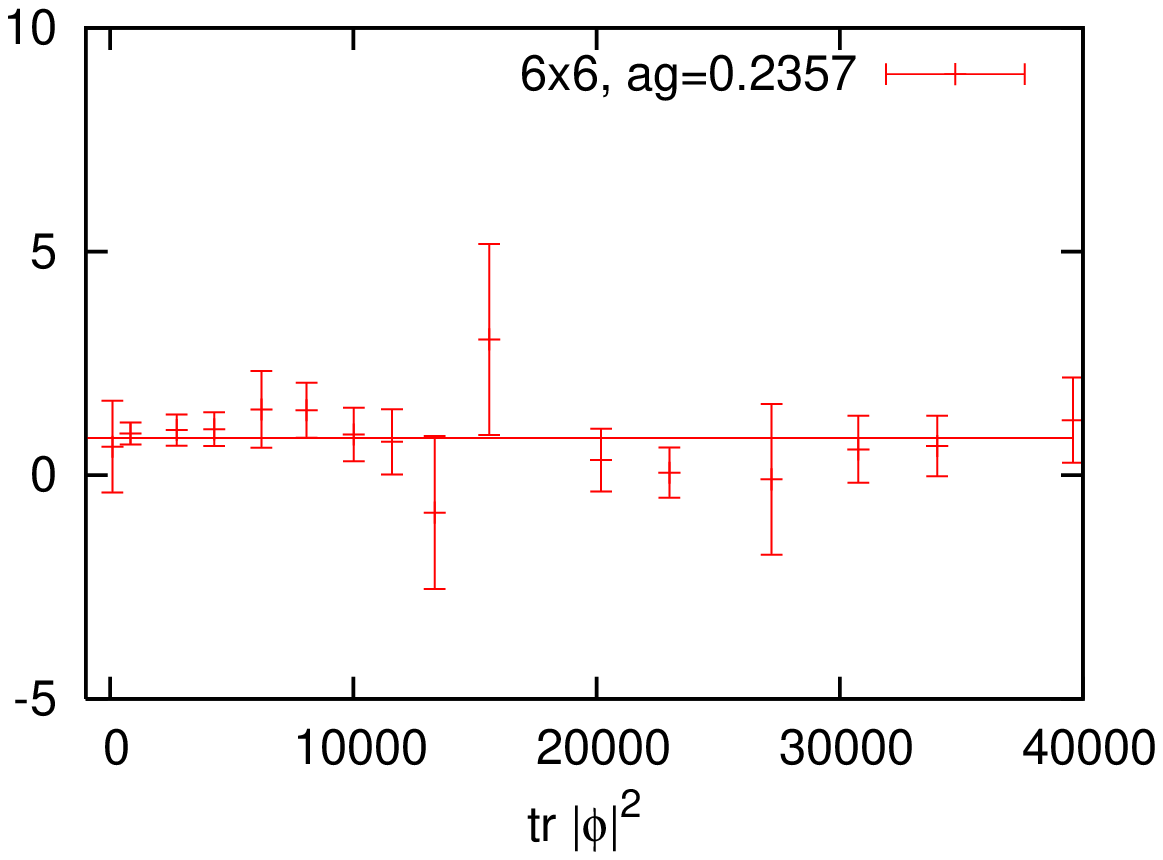}  
  \caption{Scalar norm dependence of the hamiltonian density at
  fixed lattice spacing.  No dependence can be found.}
  \label{fig:h_vs_s}
 \end{minipage}
\end{figure}

A potential danger comes from the non-compact flat direction of
scalar fields.  
The current lattice model as well as the target theory in the continuum
has classical flat directions.
Figure~\ref{fig:scalar} shows in fact the scalar fields 
are not stabilized at the origin of the potential.  
It rather goes far away over the cut off scale.\footnote{
Effects of the large scalar fields will be discussed in \cite{restore}.
}
We regard this fact as an evidence that we have actually integrated over the
non-compact configuration space of the scalar fields. 
In fact the quantity of interest, the Hamiltonian density, 
does not depends on the 
norm of the scalars (Fig.~\ref{fig:h_vs_s}).

\begin{table}
 \hfil
 \begin{tabular}{c|c|cccccc} \hline
 \multicolumn{2}{c|}{} & \multicolumn{6}{c}{$N_T/N_S$} \\
 \hline
 $N_S$ & $ag$ & 0.25 & 0.5 & 1 & 1.5 & 2  & 3 \\
 \hline
  6   & 0.2357  & ---  &   500 & 1,700 & 1,300 &  1,000 & 1,100  \\
  8   & 0.1768  & ---  &   500 & 1,100 & 1,100 &    280 &   700  \\
  10  & 0.1414  & ---  &   --- &   --- &   --- &    --- &   175  \\
  12  & 0.1179  &  20  &   600 &   110 &   450 &    500 &   ---  \\
  16  & 0.08839 &  10  &   --- &   --- &   ---    & --- &   ---  \\
  20  & 0.07071 &  20  &   --- &   --- &   ---    & --- &   ---  \\
 \hline
 \end{tabular}
 \caption{Numbers of independent configurations after binning for
 $N_T \times N_S$ lattice, $N_T$ refers temporal direction $N_S$ refers
 spacial direction.}
 \label{table:configurations}
\end{table}

\section{Conclusion and Discussion}
We carried out the RHMC simulation for two-dimensional $\mathcal{N}=(2,2)$ 
super Yang-Mills 
based on Sugino model, which exactly keeps one scalar supercharge.
Using the simulation, we observed the ground state energy
which is useful to check the dynamical SUSY breaking.
Compared with the result in \cite{prd,ptp}, which 
did not utilize the dynamical fermion but the fermion
effects were reweighted, the current simulation drastically reduced
the error.  
An extension to couple the matter multiplet based on \cite{Sugino:2008yp}
will be an interesting application.

Before giving the conclusive result with respect to SUSY breaking
using this simulation,  
we should check whether the current lattice model actually describes the
target continuum theory.
We should check the restoration of the other three supercharges
explicitly.
It is no longer an assumption based on the 
perturbative discussion but the current simulation with dynamical fermion
allows us to give an explicit numerical check \cite{restore}.

\acknowledgments{
The author would like to thank F.~Sugino and H. Suzuki for useful discussion.
He also thanks H.~Kawai, M.~Hanada, H.~Matsufuru and K.~Murakami.  
This work is supported by the Special Postdoctoral Researchers Program
 in RIKEN.  
The numerical simulation was done using Riken Super Combined Cluster(RSCC). 
The simulation code in C++ was developed based on \cite{DiPierro:2000bd}.
}

\end{document}